\documentclass[11pt,letter]{article}

\usepackage{graphicx}
\usepackage{amsmath}
\usepackage{amsfonts}
\usepackage{amssymb}
\usepackage{natbib}
\usepackage{subfigure}

\newcommand{\re}{{\mathbb{R}}}
\newcommand{\D}{\mathcal{D}}
\newcommand{\dd}{{\mathrm{d}}}
\newcommand{\qq}{{\mathrm{q}}}

\newcommand{\ipr}[2]{\langle #1, #2 \rangle}
\newcommand{\vc}[1]{{\boldsymbol{#1}}}

\title{\bf $L^1$-Optimal Splines for Outlier Rejection%
\thanks{This research is supported in part by the JSPS Grant-in-Aid for Scientific Research (C) No.~24560543}
}
\author{Masaaki Nagahara$^{1\dagger}$ and Clyde F. Martin$^2$\\
$^1$Kyoto University, Kyoto JAPAN\\
$^\dagger$Corresponding author, e-mail: \texttt{nagahara@ieee.org}\\
$^2$Texas Tech University, Texas, USA\\
}
\date{}
\begin{document}
\maketitle
\thispagestyle{empty}

\begin{abstract}
In this article, we consider control theoretic splines 
with $L^1$ optimization for rejecting outliers in data. 
Control theoretic splines are either interpolating or smoothing splines, 
depending on a cost function
with a constraint defined by linear differential equations. 
Control theoretic splines are effective for Gaussian noise in data 
since the estimation is based on $L^2$ optimization. 
However, in practice, there may be outliers in data, 
which may occur with vanishingly small probability under the Gaussian assumption 
of noise,
to which $L^2$-optimized spline regression may be very sensitive. 
To achieve robustness against outliers, we propose to use $L^1$ optimality, 
which is also used in support vector regression.
A numerical example shows the effectiveness of the proposed method.
\end{abstract}

\section{Introduction}
{\em Control theoretic spline} is generalization
of the smoothing spline proposed in \cite{KimWah71,Wah},
using control theoretic ideas,
by which the spline curve is determined by
the output of a linear control system.
Control theoretic splines give a richer class of smoothing curves
relative to polynomial curves.
They have been proved to be useful for trajectory planning in \cite{EgeMar01},
mobile robots in \cite{TakMar04}, 
contour modeling of images in \cite{KanEgeFujTakMar08},
probability distribution estimation in \cite{Cha10}, to name a few. 
For more
applications and a rather complete theory of control theoretic
splines, see \cite{EgeMar}.

Control theoretic splines are effective for reducing Gaussian noise
in data since the estimation is based on $L^2$ optimization.
This means that the noise distribution is assumed to decay very rapidly
as the amplitude increases ($\propto e^{-|x|^2/2}$).
However, in practice, there may exist {\em outliers}
in data, which may occur with vanishingly small probability
under the assumption of the Gaussian distribution of noise.
To such noise, $L^2$-optimized spline regression may
be very sensitive.

Instead, we adopt \emph{$L^1$} optimality for regression
to achieve robustness against outliers.
That is, we assume \emph{Laplacian distribution} for noise,
which is distributed much more slowly ($\propto e^{-|x|}$)
than Gaussian tails.
This is related to
{\em support vector regression} (SVR) 
(see e.g. \cite{SchSmo}),
which can be reduced to convex optimization that can be
efficiently solved by numerical optimization
(see e.g. \cite{BoyVan}).


\section{$L^1$-Optimal Splines}
Consider the following linear system $P$:
\begin{equation}
 P:\left\{\begin{split}
 \dot{\vc{x}}(t) &= A\vc{x}(t) + \vc{b}u(t), & \\y(t) &= \vc{c}^\top \vc{x}(t), &t\in[0,\infty).
 \end{split}\right.
 \label{eq:system}
\end{equation}
where $A\in\re^{n\times n}$, $\vc{b},\vc{c}\in\re^{n}$,
$\vc{x}(0)=\vc{0}\in\re^{n}$.
We assume $(A,\vc{b})$ is controllable and $(\vc{c}^\top,A)$ is observable.
For this system, suppose that we are given data
$\D := \{(t_1, y_1), (t_2, y_1), \ldots (t_N, y_N)\}$,
where $t_1,\ldots,t_N$ are sampling instants which satisfy
$0<t_1<t_2<\cdots<t_N=:T$.
The objective here is to find the control input $u(t)$, $t\in[0,T]$ for the system \eqref{eq:system}
such that
$y(t_i)\approx y_i$ for $i=1,\ldots,N$.
For this purpose, the following \emph{quadratic} cost function has been introduced in 
\cite{SunEgeMar00,EgeMar}:
\begin{equation}
 J_\qq(u) := \lambda \int_0^T |u(t)|^2 \dd t + \sum_{i=1}^N w_i (y(t_i) - y_i)^2,
 \label{eq:J2}
\end{equation}
where 
$\lambda>0$ is the regularization parameter that specifies the tradeoff between
the smoothness of $u$ defined in the first term of \eqref{eq:J2}
and the minimization of the empirical risk in the second term.
Also, $w_i>0$ is a weight for $i$-th empirical risk.
The optimal control $u=u_\qq^\ast$ that minimizes $J_\qq(u)$ is given by
\cite{SunEgeMar00,EgeMar} as
\begin{equation}
 u_\qq^\ast(t) = \sum_{i=1}^N \theta_{\qq}^\ast[i] g_{i}(t),
 \label{eq:optimal_u2}
\end{equation}
where $g_{i}(t)$ is defined by
\begin{equation}
 g_{i}(t) := \begin{cases} 
	\vc{c}^\top e^{A(t_i-t)}\vc{b},
	&\quad \text{if}~~t_i>t,\\ 
	0,&\quad \text{if}~~t_i\leq t,
 \end{cases}
 \label{eq:gi}
\end{equation}
and the optimal coefficients $\theta_\qq[1],\dots,\theta_\qq[N]$ are given by
given by
\begin{equation}
 \vc{\theta}_\qq^\ast:=\bigl[\theta_{\qq}^\ast[1],\ldots,\theta_{\qq}^\ast[N]\bigr]^\top=(\lambda W^{-1} + G)^{-1}\vc{y},
 \label{eq:L2optimal}
\end{equation}
where
$\vc{y}:=[y_1,\ldots, y_N]^\top$.
The matrix $G$ in \eqref{eq:L2optimal}
is the Grammian defined by $G_{ij}=\ipr{g_i}{g_j}$, $i,j=1,\ldots,N$.

In \eqref{eq:J2}, the empirical risk (the second term)
is measured by $L^2$ norm.
This is based on the assumption that
the noise added to the data is Gaussian.
However, there may exist outliers
in data, which may be ignored
under the Gaussian assumption of noise.
To such outliers, the regression may be very sensitive.

To overcome this, we introduce the following
distribution function instead of Gaussian,
called {\em $\epsilon$-insensitive function} (see \cite{SchSmo}):
\begin{equation}
 p(\xi) = \frac{1}{2(1+\epsilon)}\exp(-|\xi|_\epsilon),
 \label{eq:e-insensitive}
\end{equation}
where $\epsilon>0$ is a fixed parameter and
$|\xi|_\epsilon := \max(|\xi|-\epsilon, 0)$.
The distribution is an "insensitive" version of the Laplace distribution,
which is given by setting $\epsilon=0$.

The distribution \eqref{eq:e-insensitive}
has heavier tails than the Gaussian distribution,
and hence leads to more robust regression against outliers.
Assuming this distribution, we introduce the following cost function:
\begin{equation}
 J := \frac{1}{2}\|\vc{\theta}\|_2^2 + C \sum_{i=1}^N \bigl|\vc{\theta}^\top \vc{\varphi}_i - y_i\bigr|_\epsilon,
 \label{eq:J1}
\end{equation}
where $C>0$ is the regularization parameter
and $\vc{\varphi}_i := \left[G_{i1},G_{i2},\ldots,G_{iN}\right]^\top$.

The optimization above can be effectively solved by employing the method of support vector regression (see \cite{SchSmo}).
That is, the optimization is reduced to the following convex optimization:
\[
 \begin{split}
  \text{minimize}~~ &\frac{1}{2}\|\vc{\theta}\|_2^2 + C\sum_{i=1}^N \left(\xi_i + \hat{\xi}_i\right)\\
  \text{subject to}~~ 
    &y_i \leq \vc{\theta}^\top \vc{\varphi}_i + \epsilon + \xi_i,~
  y_i \geq \vc{\theta}^\top \vc{\varphi}_i - \epsilon - \hat{\xi}_i,~
  \xi_i \geq 0,~
  \hat{\xi}_i\geq 0.
 \end{split}
\]
\section{Example}
\begin{figure}[tbp]
 \centering
 \includegraphics[width=0.9\linewidth]{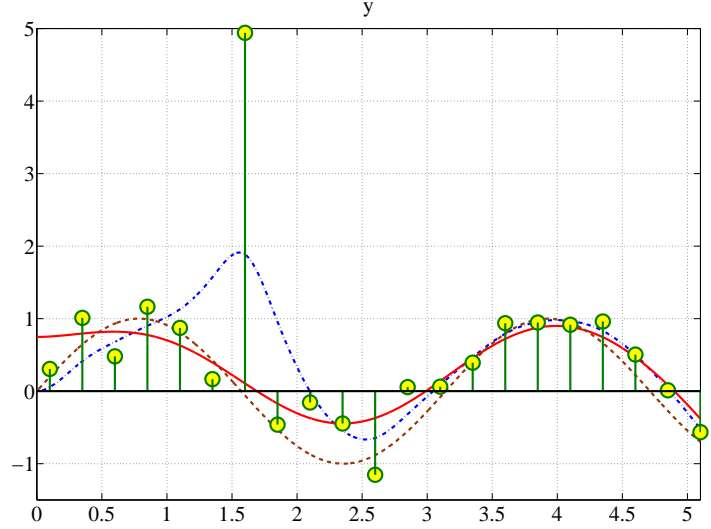}
 \caption{Original curve $y(t)=\sin 2t$ (dashed line), sampled data $\D$ (circles),
 reconstructed curves via $L^1$ optimization (solid line) and via $L^2$ optimization (dash-dotted line).}
 \label{fig:data}
\end{figure}

We here show an example.
Let us consider a linear system \eqref{eq:system} with
\[
  A = \begin{bmatrix}0&1\\1&0\end{bmatrix},\quad \vc{b} = \begin{bmatrix}1\\0\end{bmatrix},\quad
  \vc{c}^\top = \begin{bmatrix}0&1\end{bmatrix}.
\]
Note that this system has its transfer function as 
$P(s)=\frac{1}{(s+1)(s-1)}$.
The sampling instants are taken by
$t_i=0.1+(i-1)\times 0.25$, 
$i=1,2,\ldots,21$,
and the data $y_i$ is generated by
$y_i=\sin 2t_i + e_i$,
where $e_1,\ldots,e_{21}$ are noise and including an outlier at $t_7=1.6$
as shown in Fig.~\ref{fig:data}.
\begin{figure}
 \centering
 \includegraphics[width=0.9\linewidth]{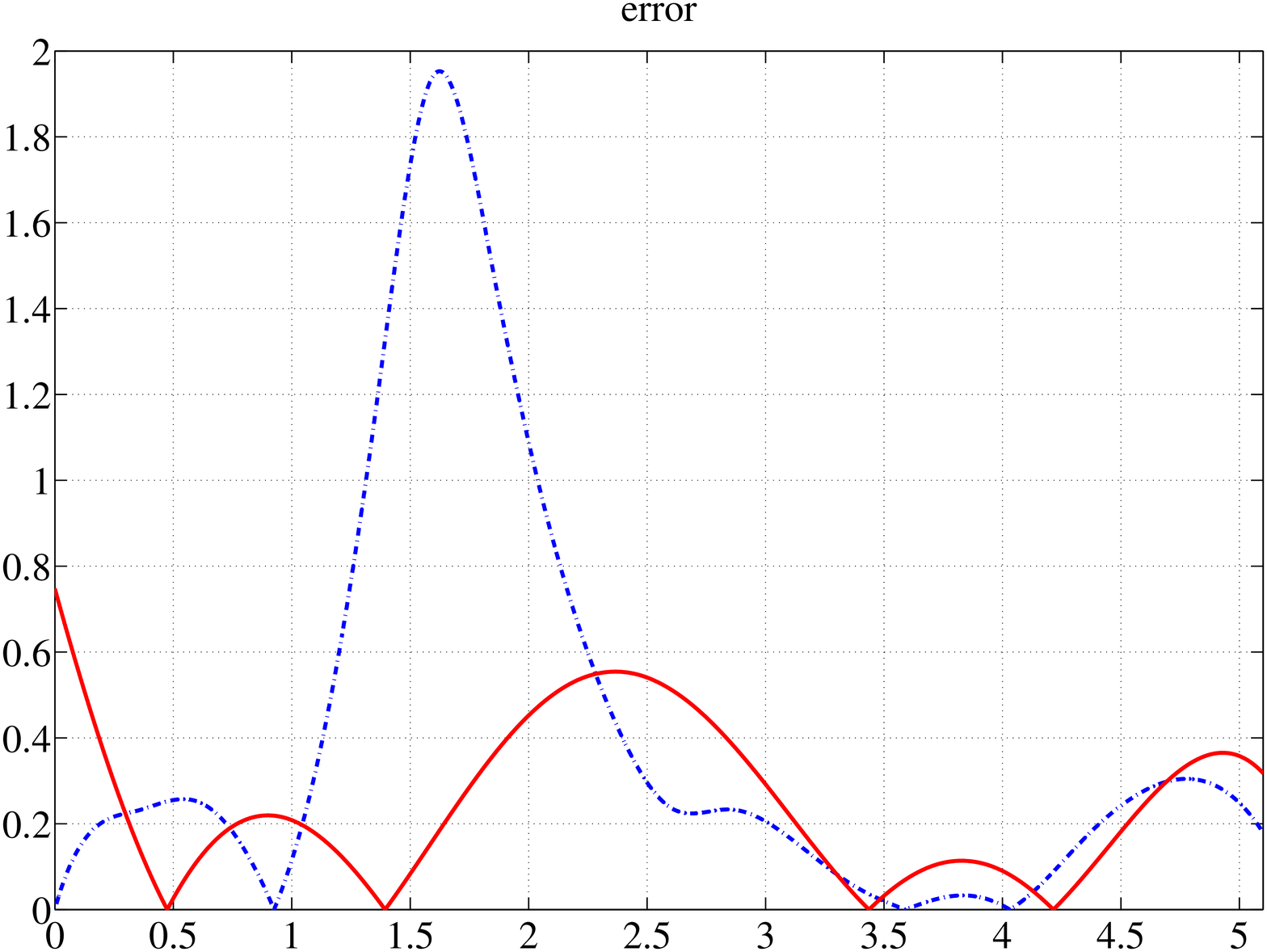}
 \caption{Regression error via $L^1$ optimization (solid line) and via $L^2$ optimization (dash-dotted line).}
 \label{fig:error}
\end{figure}

With these data, we compute two kinds of control;
the proposed control $u^\ast$ with $\vc{\theta}=\vc{\theta}^\ast$ minimizing $J$ in \eqref{eq:J1},
and the conventional $L^2$-optimal control $u_\qq^\ast$ minimizing $J_\qq$ in \eqref{eq:J2}.
The regression result is shown in Fig.~\ref{fig:data}.
Fig.~\ref{fig:error} also shows the regression error.
We can see that the conventional $L^2$-optimal regression shows
high sensitivity to the outlier and around there the regression error becomes very large.
On the other hand, the result by the proposed $L^1$ optimization shows much more robust regression.

\section{Conclusion}

In this article, we have proposed outlier rejection for control theoretic splines
based on $L^1$ optimization. 
While conventional $L^2$-based splines are effective for Gaussian noise,
they are very sensitive to outliers.
To achieve robustness against them, we have propose to adopt
$L^1$ optimality in the regression.
A numerical example has shown the effectiveness of our method.
Future works include to design constrained splines as discussed in \cite{NagMar13},
and to investigate the sparsity property of $L^1$-optimal splines
as in \cite{NagQue11,NagMatHay12,NagQueOst12b}.


\end{document}